\documentclass[pdflatex,sn-mathphys]{sn-jnl}

\usepackage{amsmath}
\usepackage{amssymb}
\usepackage{graphicx}

\usepackage{algorithm}
\usepackage{algpseudocode}
\usepackage{xurl}

\algnewcommand{\LineComment}[1]{\State \parbox[t]{\dimexpr\linewidth-\leftmargin}{\raggedright{\(\triangleright\) #1}\strut}}

\usepackage{longtable}

\jyear{2022}%

\theoremstyle{thmstyleone}%
\theoremstyle{thmstyletwo}%

\theoremstyle{thmstylethree}%

\begin{document}

\title[\small{Fuel savings through missed approach maneuvers based on aircraft reinjection}]{Fuel savings through missed approach maneuvers based on aircraft reinjection}


\author[1]{\fnm{María} \sur{Carmona}}\email{Maria.Carmona@uclm.es}
\author*[1]{\fnm{Rafael} \sur{Casado}}\email{Rafael.Casado@uclm.es}
\author[1]{\fnm{Aurelio} \sur{Bermúdez}}\email{Aurelio.Bermudez@uclm.es}
\author[2]{\fnm{Miguel} \sur{Pérez-Francisco}}\email{mperez@uji.es}
\author[3]{\fnm{Pablo} \sur{Boronat}}\email{boronat@lsi.uji.es}
\author[4]{\fnm{Carlos} \sur{T. Calafate}}\email{calafate@disca.upv.es}

\affil*[1]{
\orgdiv{Albacete Research Institute of Informatics}, \orgname{University of Castilla-La Mancha}, \orgaddress{\street{Campus Universitario s/n}, \city{Albacete}, \postcode{02071}, \country{Spain}}}

\affil[2]{\orgdiv{Computer Science and Engineering Department}, \orgname{Universitat Jaume I (UJI)}, \orgaddress{\street{Av. Sos Baynat S/N}, \city{Castell\'o de la Plana}, \postcode{12071}, \country{Spain}}}

\affil[3]{
\orgdiv{Computer Languages and Systems Department}, \orgname{Universitat Jaume I (UJI)}, \orgaddress{\street{Av. Sos Baynat S/N}, \city{Castell\'o de la Plana}, \postcode{12071}, \country{Spain}}}

\affil[4]{
\orgdiv{Department of Computer Engineering (DISCA)}, \orgname{Universitat Politècnica de València}, \orgaddress{\street{Camino de Vera, S/N}, \city{Valencia}, \postcode{46022}, \country{Spain}}}





\abstract{
Humanity is facing global challenges related to climate change, along with an energetic crisis that urgently requires optimizing any process or system able to improve global economic conditions. Taking fuel as an example, its prices are among the highest increases detected in goods of general use. In this sense, initiatives to mitigate fuel consumption are both welcome and necessary. In the aerial transportation area, fuel costs are critical to the economic viability of companies, and so urgent measures should be adopted to avoid any unnecessary increase of operational costs. In this work we study the costs involved in a standard procedure following a missed approach. In addition, we study the improvements achieved with a fast reinjection scheme proposed in a prior work. Experimental results show that, for a standard A320 aircraft, fuel savings ranging from 55\% to 90\% can be achieved through our proposed method.}

\keywords{precision approach; missed approach; fuel savings; base aircraft data}



\maketitle
\section*{Nomenclature}

\begingroup
\setlength{\tabcolsep}{10pt} 
\renewcommand{\arraystretch}{1.5} 

\begin{longtable}{ l l }

 $D$  &  aerodynamic drag force ($N$) \\
 $C_D$  &  drag coefficient \\
 $C_{D_0}$  &  parasitic drag coefficient \\
 $C_{D_2}$  &  induced drag coefficient \\
 $C_{f1}$  &  1st trust specific fuel consumption coefficient $(kg/(min \times kN))$ \\
 $C_{f2}$  &  2nd trust specific fuel consumption coefficient $(kt)$ \\
 $C_{fcr}$  &  cruise fuel consumption correction coefficient \\
 $\gamma$  &  path angle $(rad)$ \\
 $\rho$  &  local air density $(kg/m^3)$ \\
 $C_L$  &  lift coefficient \\
 $F$  &  fuel flow $(kg/min)$ \\
 $g$  &  gravitational acceleration $(m/s^2)$ \\
 $h$  &  geodetic altitude $(m)$ \\
 $L$  &  aerodynamic lift force $(N)$ \\
 $m$  &  aircraft mass $(kg)$ \\
 $S$  &  reference wing surface area $(m^2)$ \\
 $T$  &  engine thrust force $(N)$ \\
 $V$  &  true airspeed $(m/s)$ \\
 $W$  &  aircraft weight $(N)$ \\
 $\eta$  &  trust specific fuel flow $(kg/(min \times kN))$ \\
 $\mathbf{a}$  &  aircraft state vector $\begin{bmatrix}\mathbf{x}&\mathbf{w}\end{bmatrix}$ \\
 $\mathbf{g}$  &  ghost aircraft state vector $\begin{bmatrix}\mathbf{x}&\mathbf{w}\end{bmatrix}$ \\
 $\mathbb{A}$  &  set of aircraft state vectors \\
 $\mathbf{x}$  &  aircraft pose $\begin{bmatrix}\mathbf{p}&\psi\end{bmatrix}$ \\
 $\mathbb{X}$  &  set of aircraft poses \\
 $\mathbf{p}$  &  aircraft 3D position $\begin{bmatrix}x&y&z\end{bmatrix}$ \\
 $\mathbb{P}^{3D}$  &  set of aircraft 3D positions \\
 $x$  &  x-coordinate $(m)$ \\
 $y$  &  y-coordinate $(m)$ \\
 $z$  &  z-coordinate $(m)$ \\
 $\psi$  &  aircraft heading (rad) \\
 $\mathbf{w}$  &  waypoint $\begin{bmatrix}\mathbf{p}&s\end{bmatrix}$ \\
 $\mathbb{W}$  &  set of waypoints \\
 $s$  &  waypoint horizontal speed $(m/s)$ \\
 $\mathbb{S}$  &  set of waypoint horizontal speeds \\
 $M$  &  approach waypoint sequence \\
 $T_s$  &  aircraft spacing $(s)$ \\
 $T_1$  &  threshold time (for gap search) $(s)$ \\
 $T_{air}$  &  air temperature $(K)$ \\
 $p$  &  air pressure $(Pa)$ \\
 $R$  &  air constant $(m^2/Ks^2)$ \\
\end{longtable}
\endgroup

\section{Introduction}


Nowadays we face important problems related to the very high number of aircraft in our skies, which generate air congestion in the main airports worldwide. In fact, by 2040 the air traffic in Europe is expected to grow over 16.2 million flights \citep{Eurocontrol_Challenges_2040}, which will increase pressure on airport capacity, and will reduce the margin of airports for handling contingency. Such congestion issues should be dealt with efficiently, as they are prone to cause an excess of noise, fuel consumption and pollutant emissions.

Conventional aircraft emit different pollutants such as carbon, nitrogen, or sulphur oxides, but also unburnt fuel and other particulates.
In particular, CO$_2$ emissions are usually taken as a reference metric to determine the overheads associated to different phases of the flight, so as to allow determining which phases have more margin for improvement.

\begin{figure}
\centering
\includegraphics[width=0.9\textwidth]{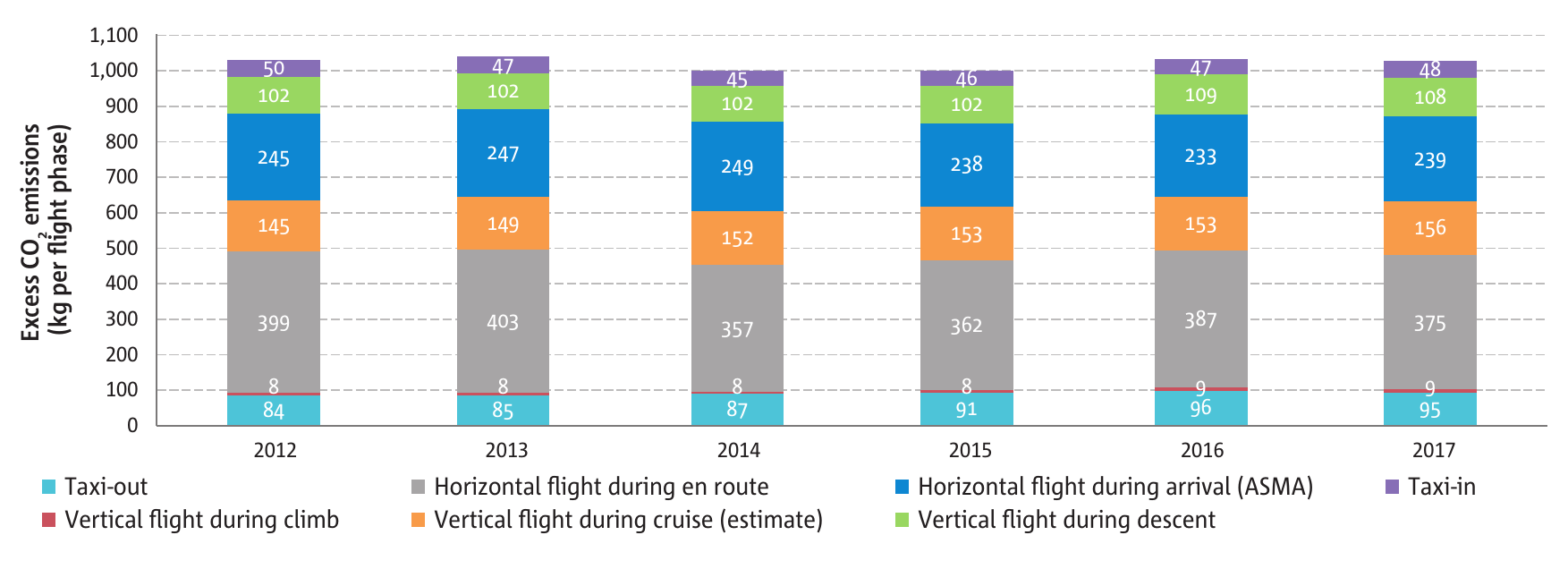}
\caption{Excess $CO_2$ emissions for an average flight in Europe in 2017 (source: EASA \citep{EASA_environmental_report}).}
\label{fig:EASA}
\end{figure}

According to \citep{EASA_environmental_report}, when an\-alysing the trajectories of all European flights in 2017, there is a 5.8\% overhead (when compared against their unimpeded trajectories) regarding CO$_2$ emissions. This means that there is still room to reduce pollutant emissions (see Figure \ref{fig:EASA}). The 2018 European ATM Master Plan \citep{ATM_Master_Plan} ambition is to arrive to a 3.2\% reduction of the time of flights, and to a 2.3\% CO$_2$ emissions reduction by 2035.
As early as 2006, Lu and Morrell \cite{2006LuMorrell} proposed a pollution and noise model for airports from an economical perspective, as the intention of many countries is to restrict the pollution regulations. This model is focused on landing and take-off operations.

Figure \ref{fig:EASA} highlights that, despite all traffic management operations already benefit from highly optimized procedures to minimize the costs involved, there is still room for improvement, especially for the take-off and landing processes  \citep{2019RodriguezAdensoGonzalez-ImprovingAircraftApproach,Murrieta-Mendoza2016}. In particular, it is well known that one of the procedures with margin for optimization is the handling of missed approaches, especially in large airports.

The conventional missed approach procedure consists of restarting the maneuver from the beginning. In \citep{Casado2021}, an alternative procedure is proposed and evaluated. It basically redirects the affected aircraft towards an existing gap in the approach flow. The Aircraft Reinjection System (ARS) detects the existence of that gap, estimates its future position, and determines a new set of waypoints that define a route to quickly reinject the aircraft in the flow. These waypoints are provided to the aircrew by the air controllers. As a result, the aircraft is fully integrated as standard traffic on the approach flow. At the same time, the new route maintains International Civil Aviation Organization (ICAO) separation standards \citep{ICAO2016}.

In this paper we make a deep analysis on how the proposed ARS solution is able to benefit both the air company and the environment by reducing the amount of fuel spent on the landing maneuvers whenever a problem causes an aircraft reinjection to take place. 
Experiments performed using our landing simulation tool show that on a standard A320 aircraft, fuel savings ranging from 55\% to 90\% can be achieved. 


The remainder of this paper is structured as follows. In the next section we present some related works on this topic. In Section \ref{sec:missedapproaches} we describe the conventional missed approach procedure, and present an overview of our proposed solution (ARS). Then, in Section \ref{sec:BADA} we detail the Base of Aircraft Data version 3 (BADA3) model that provides the performance (including fuel consumption) of an aircraft during take-off and landing when following standard procedures. After that, Section \ref{sec:perf_eval} presents the simulation methodology and the results obtained in this work. Finally, the main conclusions and future works are drawn in section \ref{sec:conclusions}. 

\section{Related Work}
\label{sec:related}

In this work we use the Base of Aircraft Data (BADA) \citep{Eurocontrol2022a, Nuic2010} to estimate the amount of fuel burnt during the missed approach maneuver. BADA is an aircraft performance model developed by the European Organization for the Safety or Air Navigation (Eurocontrol) that can be used to generate aircraft trajectories and to estimate fuel consumption and emissions in all flight phases. The accuracy of the BADA model has been validated by several studies \citep{Poles2010, HARADA2013}, and it has been used as a baseline reference in a large amount of related works. Other aviation authorities have designed tools to estimate burnt fuel and emissions. This is the case of the Aviation Environmental Design Tool (AEDT) \citep{FederalAviationAdministrationFAA2022}, developed by the Federal Aviation Administration (FAA) to estimate fuel consumption, emissions, noise, and air quality consequences, and the Carbon Emission Calculator \citep{ICAO2022a}, provided by ICAO to calculate the carbon dioxide emissions from air travel.

Additionally, several fuel burn prediction models have been developed by researchers. Some of them are based on complex physics that consider the aircraft characteristics, its trajectory, the atmospheric conditions, etc. A detailed review of these models is provided in \citep{Yanto2018}. Most of them apply the so called “Breguet Range Equation” \citep{Breguet1923}, which considers the rate of an aircraft’s mass change during its flight, being unsuitable for the descent segment. In other cases, fuel burn models are based on empirical data provided by Flight Data Recorders (FDR) \citep{Chati2014, Chati2018}, Quick Access Recorders \citep{Zhang2019,Ye2019}, radars \citep{Liu2018}, or databases provided by aviation organizations \citep{Yanto2018, 2019RodriguezAdensoGonzalez-ImprovingAircraftApproach}. Within the second category, many recent works propose to apply machine learning techniques, such as neural networks (ANN) \citep{Hong2018, Baumann2020}, decision trees \citep{Baumann2020}, and support vector machines (SVM) \citep{Wang2015,Haifeng2015}, to estimate fuel consumption from the measured data.

Some of these works focus on the Landing and Take-Off cycle (LTO), which comprises all the flight phases which are below 3000 ft Above Ground Level (AGL). These studies point out the relevance of these flight phases in terms of pollution, noise, and fuel consumption, which particularly affect the surroundings of airports.
For example, \citep{Chati2018} develops statistical models for aircraft fuel burn in the climb out and approach phases. In \citep{Sahin2018} the BADA model is used to estimate the fuel required during the descent phase. Models which optimize the approach maneuver, such as the one proposed in \citep{2014Salah-Environmental_impact}, show that there are still improvements that can be applied respecting actual safety regulations. These proposals are compatible with our reinjection method in case of missed approaches.

Regarding the missed approach procedure, authors in \citep{Murrieta-Mendoza2016} addressed the problem of computing the fuel consumption and the corresponding pollution emissions generated during this maneuver. The basics for these calculations were the data provided by the European Environment Agency \citep{EuropeanEnvironmentAgencyEEA2022}. Results were compared with a successful approach. As expected, authors observed that the cost of the missed approach is highly dependent on the vector provided by the ATC to exit the holding pattern, and to intercept the approach path. Similar computations were previously performed in \citep{Dancila2013} for the B737-400 aircraft.

Our work differs from the former ones as we specifically focus on the fuel savings achievable when adopting an optimized aircraft reinjection solution. 

\section{Improving the Missed Approach Procedure}
\label{sec:missedapproaches}

In this section we will detail how the missed approach procedure can be improved through our proposed solution. To this end we first detail how precision approaches work, and the conventional procedure adopted upon a missed approach. Next, we summarize the proposed Aircraft Reinjection System for reducing the flight path of an aircraft following a missed approach. The purpose of this section is to provide sufficient technical background on the proposed methods so as to allow a better understanding of the performance gains detailed later in section \ref{sec:perf_eval}.

\subsection{Precision Approaches and Conventional Missed Approach}

This work is focused on the approach flight phase. More specifically, we are interested in standard instrument approach procedures (or IAP) \citep{FAA2022a}. In this type of approaches, pilots are supported by several navigation aids located at or nearby the runway, and follow an instrument approach chart (or approach plate) that contains all the necessary information. An IAP comprises several segments, referred to as initial, intermediate and final segments, and may also include a missed approach segment. Each approach segment is conveniently defined by a pair of starting and ending fixes, which correspond with geographical positions.

The initial segment starts at the Initial Approach Fix (IAF), and is used by the aircraft to transit from an en-route airway to the Intermediate Fix (IF). The intermediate segment allows then descending to an intermediate altitude, and aligning the aircraft to the runway. After reaching the Final Approach Point (FAP), the aircraft starts the last phase of the approach to land. During this final segment, it counts on the support of navigation aids, being the Instrument Landing System (ILS) \citep{Moir2013} one of the most common ones. After reaching the Missed Approach Point (MAP/MAPt) or the Decision Altitude/Height (DA/H), depending on whether it is a precision approach or not, the aircraft must land or, alternatively, start the missed approach maneuver. In non-precision approaches, aircraft rely on horizontal (2D) guidance, provided, for example, by VOR (VHF Omnidirectional Range) stations. On the other hand, in precision approaches, aircraft receive both horizontal and vertical (3D) guidance, which is commonly provided by the ILS.

\begin{figure}
\centering
\includegraphics[width=0.7\textwidth]{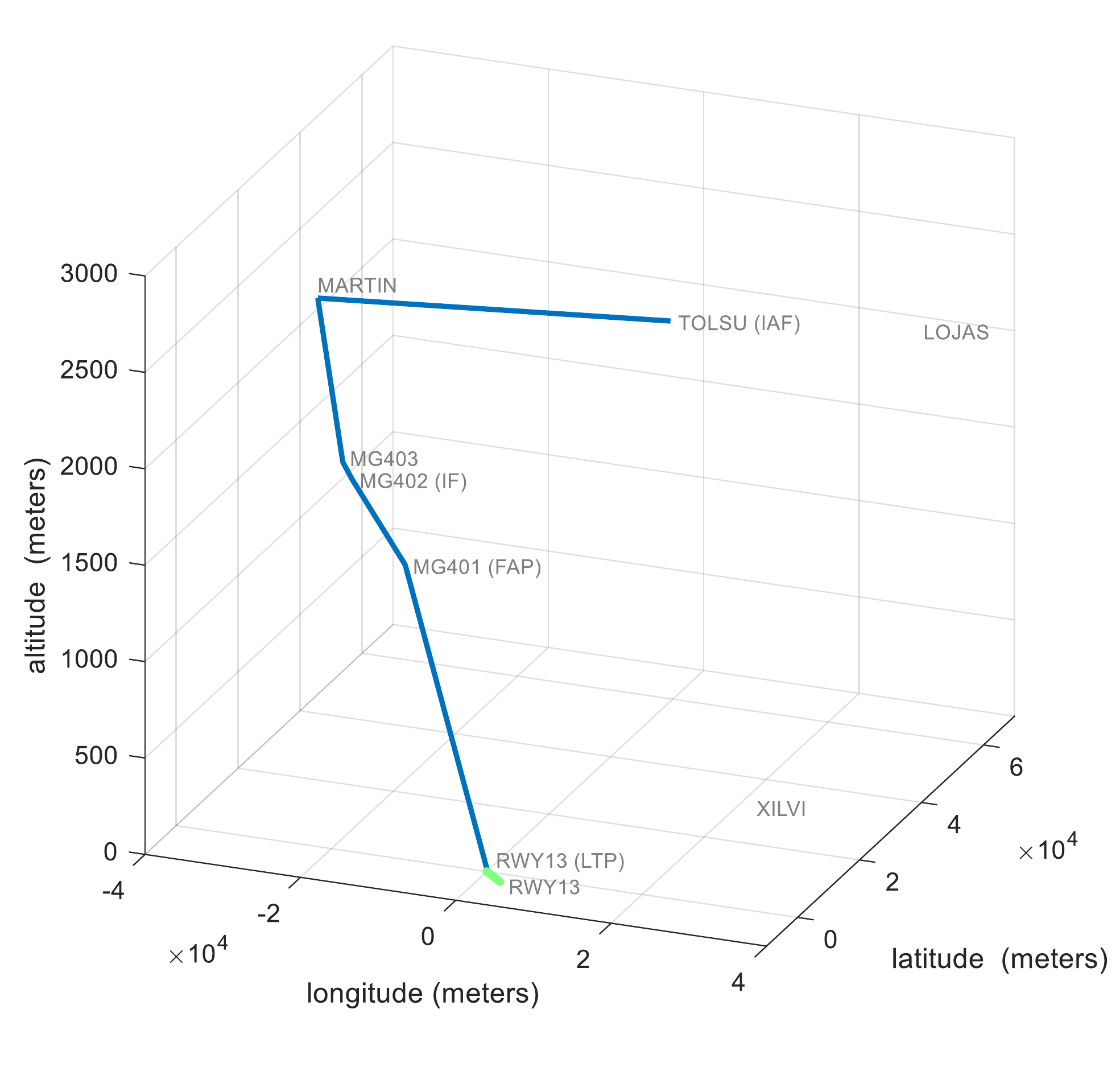}
\caption{Málaga “RWY 13” approach procedure.}
\label{fig:malaga3Dapproach}
\end{figure}

To illustrate this procedure, Figure \ref{fig:malaga3Dapproach} shows a 3D view of the approach to runway “RWY 13” at Málaga airport \citep{ICAO2022b}. Latitude and longitude are expressed in meters with respect to a reference system centered at the runway touchdown point, and altitude is expressed in meters with respect to sea level (see Figure \ref{fig:airspacemodel}). Each division on the horizontal plane represents a 20 x 20 km area. The path to be followed by each approaching aircraft is represented by a blue line, and a green line represents the runway. For the sake of simplicity, we assume that aircraft enter the airport airspace through the LOJAS waypoint. Then, they fly to TOLSU, which is the IAF according to the approach chart. After that, they proceed to MARTIN, where they turn left and progressively descend, first to the IF (MG402), and then to the FAP (MG401). From there, they cover the final approach segment to the runway. Table \ref{table:malagawaypoints} provides the position (considering the mentioned reference system) and speed associated to each of the waypoints composing this approach procedure. In this table, LTP (Landing Threshold Point) is the runway waypoint.

\begin{table}[h!]
\centering
\caption{Málaga “RWY 13” waypoints.}
\label{table:malagawaypoints}
\begin{tabular}{|l r r r r|} 
 \hline
 Waypoint & X (m) & Y (m) & Z (m) & Speed (m/s) \\ 
 \hline
 LOJAS & 32115.94 & 7950.47 & 2133.60 & 123.47 \\
 TOLSU (IAF) & 3788.66 & 49848.85 & 2133.60 & 123.47 \\
 MARTIN & -38123.21	& 41103.20 & 2133.60 & 123.47 \\
 MG403 & -29788.86 & 28279.77 & 1524 & 123.47 \\
 MG402 (IF) & -26759.25 & 23616.67 & 1524 & 82.31 \\
 MG401 (FAP) & -16175.05 & 14299.41 & 1280.16 & 82.31 \\
 RWY13 (LTP) & 55.74 & -53.08 & 15.85 & 72.02 \\
 RWY13 & 2179.44 & -2035.92 & 15.85 & 25.72 \\
 XILVI & 36907.56 & -7831.11 & 670.56 & 113.18 \\
 \hline
\end{tabular}

\end{table}

A missed approach (or go-around) procedure is the procedure to be followed by the aircraft if, due to any circumstance, an approach to land cannot be safely executed \citep{FAA2022b}. Once the pilot makes the decision of aborting the landing, he/she is expected to notify by radio to the air traffic control service (ATC) the initiation of the missed approach as soon as possible. Then, once the MAPt defined on the instrument chart has been reached, the pilot must follow the missed approach instructions indicated in the chart, or an alternative maneuver as provided by the ATC. Current missed approach procedures are based on traditional radio aids-based navigation. Most of the approach charts propose a pattern that, in a best-case scenario, reroutes the aircraft to the IAF. Note that redirecting the aircraft implies additional flying time, and a lot of workload to air controllers, who must must carry out the necessary calculations to return the aircraft to the beginning of the approach.

\begin{figure}
\centering
\includegraphics[width=0.7\textwidth]{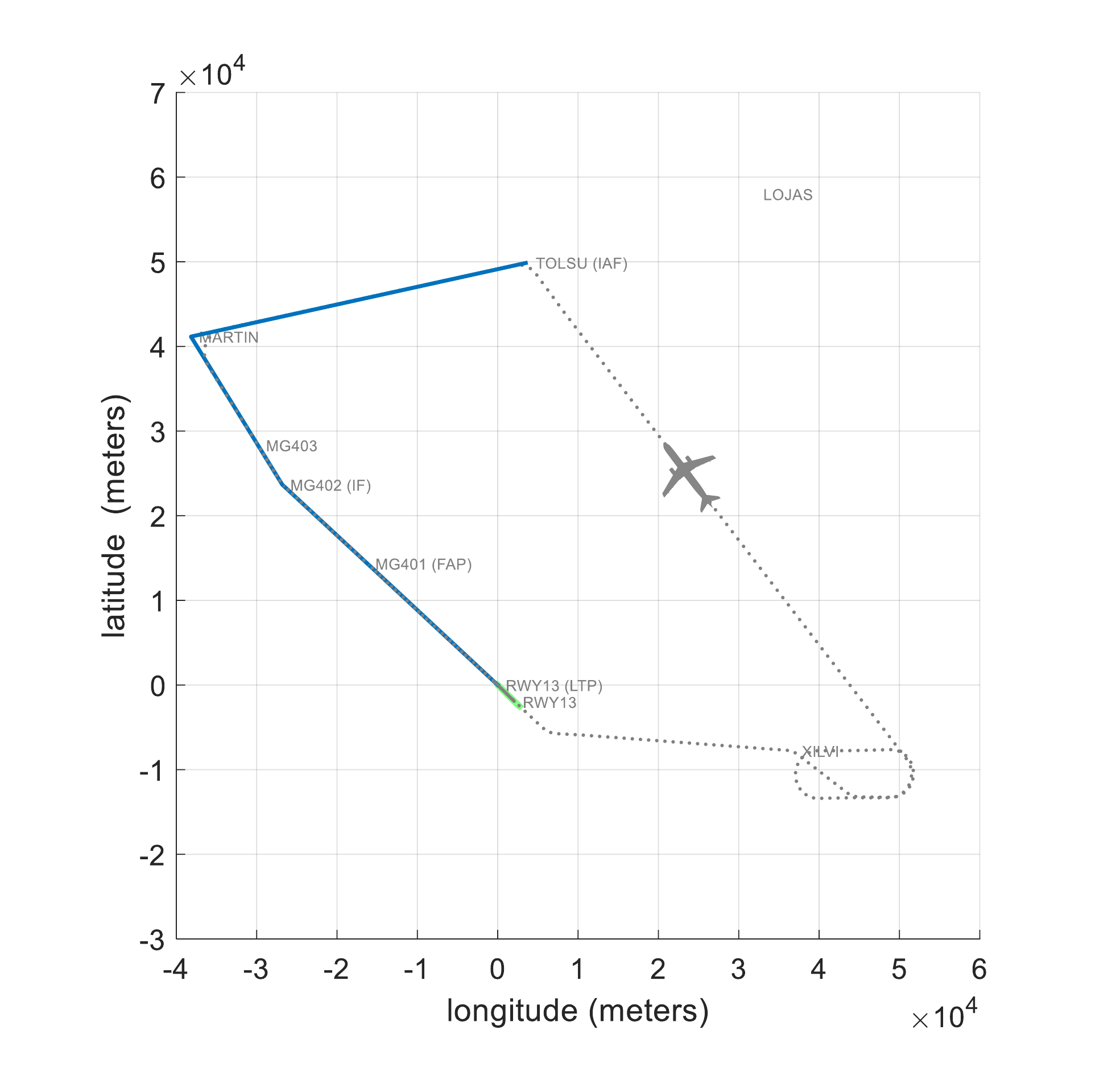}
\caption{Conventional missed approach maneuver at Málaga “RWY 13”.}
\label{fig:convmalaga}
\end{figure}

In the case of Málaga airport, the approach chart establishes that the aircraft must maintain approximately the same heading as that of the runway for about 20 NM, and then join the XILVI point. Once there, the aircrew must await ATC instructions. Figure \ref{fig:convmalaga} provides an upper view of the described missed approach maneuver, assuming that air controllers have provided clearance to proceed to TOLSU (the IAF) once the the aircraft has performed the holding pattern. Again, the blue line represents the approach path specified by the chart, and the dotted line indicates the path followed by the aircraft after aborting the landing (at the MAPt). 

\subsection{Aircraft Reinjection System (ARS)}
\label{sec:ARS}

The traditional missed approach maneuver described above introduces a high overhead to the total approach procedure. So, in a previous work \citep{Casado2021} we proposed an optimization to this procedure from a theoretical and operational perspective. We now proceed to summarize such solution.

\begin{figure}
\centering
\includegraphics[width=0.6\textwidth]{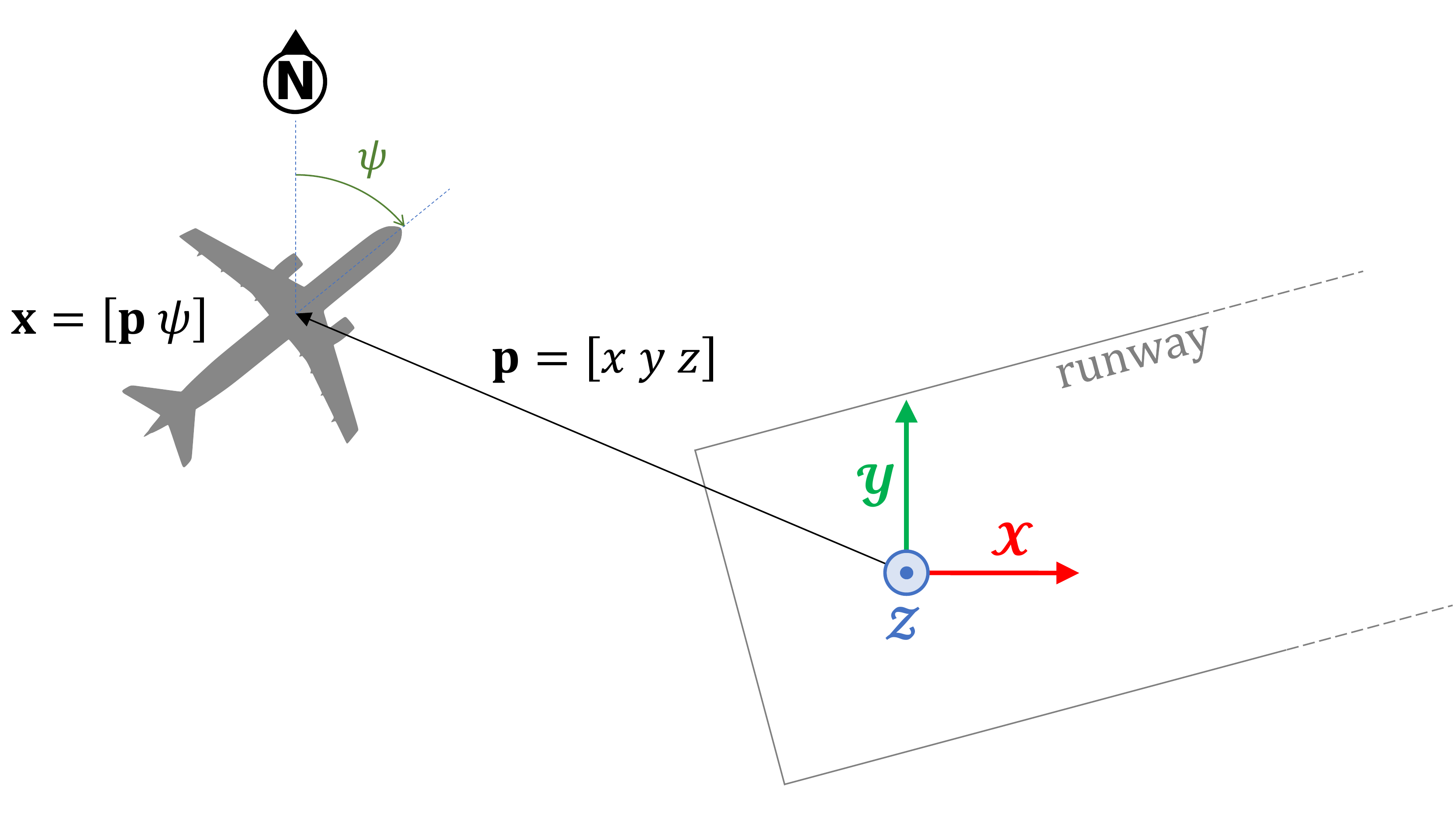}
\caption{Airspace model.}
\label{fig:airspacemodel}
\end{figure}

First, we assume that each approach is defined by a sequence of waypoints to be covered by the aircraft. By default, aircraft follow the approach defined in the instrument approach chart (introduced above), but we assume that air controllers can modify, add and remove elements in these sequences of waypoints when necessary. We formally define a waypoint as a vector $\mathbf{w}=\begin{bmatrix}\mathbf{p}&s\end{bmatrix} \in \mathbb{W}$, where $\mathbf{p}=\begin{bmatrix}x&y&z\end{bmatrix} \in \mathbb{P}^{3D}$ establishes a three-dimensional position according to the reference system shown in Figure \ref{fig:airspacemodel}, and $s \in \mathbb{S}$ determines the horizontal speed at which the aircraft must proceed to that position. We also define an approach sequence $M=\{\mathbf{w}_1,\mathbf{w}_2,...,\mathbf{w}_n\} \in \mathcal{P}(\mathbb{W})$ as an ordered list of waypoints. From this, a complete approach procedure can be defined by means of a set of alternative approach sequences ending in the same runway. This set of sequences can be conceptually structured as a tree. 

We also assume that the ATC has up-to-date information on the airspace situation at all times. More specifically, the air controller handles a set of aircraft $\mathbb{A}=\{\mathbf{a}_1, \mathbf{a}_2, \mathbf{a}_3...\}$, where each element $\mathbf{a}=\begin{bmatrix}\mathbf{x}&\mathbf{w}\end{bmatrix} \in \mathbb{A}$ represents the state vector of an approaching aircraft. The first element of the state vector ($\mathbf{x}$) is the aircraft pose. It is defined as $\mathbf{x} = \begin{bmatrix}\mathbf{p}&\psi\end{bmatrix} \in \mathbb{X}$, where  $\mathbf{p}=\begin{bmatrix}x&y&z\end{bmatrix} \in \mathbb{P}^{3D}$ is its current position, and $\psi$ is its current heading (see Figure \ref{fig:airspacemodel}). The second element ($\mathbf{w}$) indicates the waypoint to which the aircraft is currently flying.

The proposed reinjection procedure is initiated the moment when the pilot of an approaching aircraft notifies the ATC of his/her decision to execute a missed approach. The system must first determine the feasibility of executing a reinjection maneuver. This is accomplished by studying the approach flow in search of a gap between two consecutive aircraft that is large enough to allow the reinjection to take place. The condition to be met is that the time difference between these two consecutive aircraft is greater than twice the aircraft spacing ($T_s$). Note that, according to the concept of Time-Based Separation (TBS) \citep{Eurocontrol2022b}, we use times instead of distances when talking about separation between aircraft.

If the search process finds a gap satisfying the above condition, the algorithm assumes the existence of a “ghost” aircraft in the position corresponding to the minimum allowable time behind the aircraft associated to this gap. This ghost aircraft behaves exactly the same as any other aircraft executing the approach maneuver.

From this point, the ARS estimates the future position of the ghost aircraft and determines an intercepting trajectory for the aircraft missing the approach, so that both ghost and real aircraft will meet at the reinjection point. This trajectory consists of three new auxiliary waypoints that the ATC must provide to the aircraft being reinjected.

\begin{figure}
\centering
\includegraphics[width=0.8\textwidth]{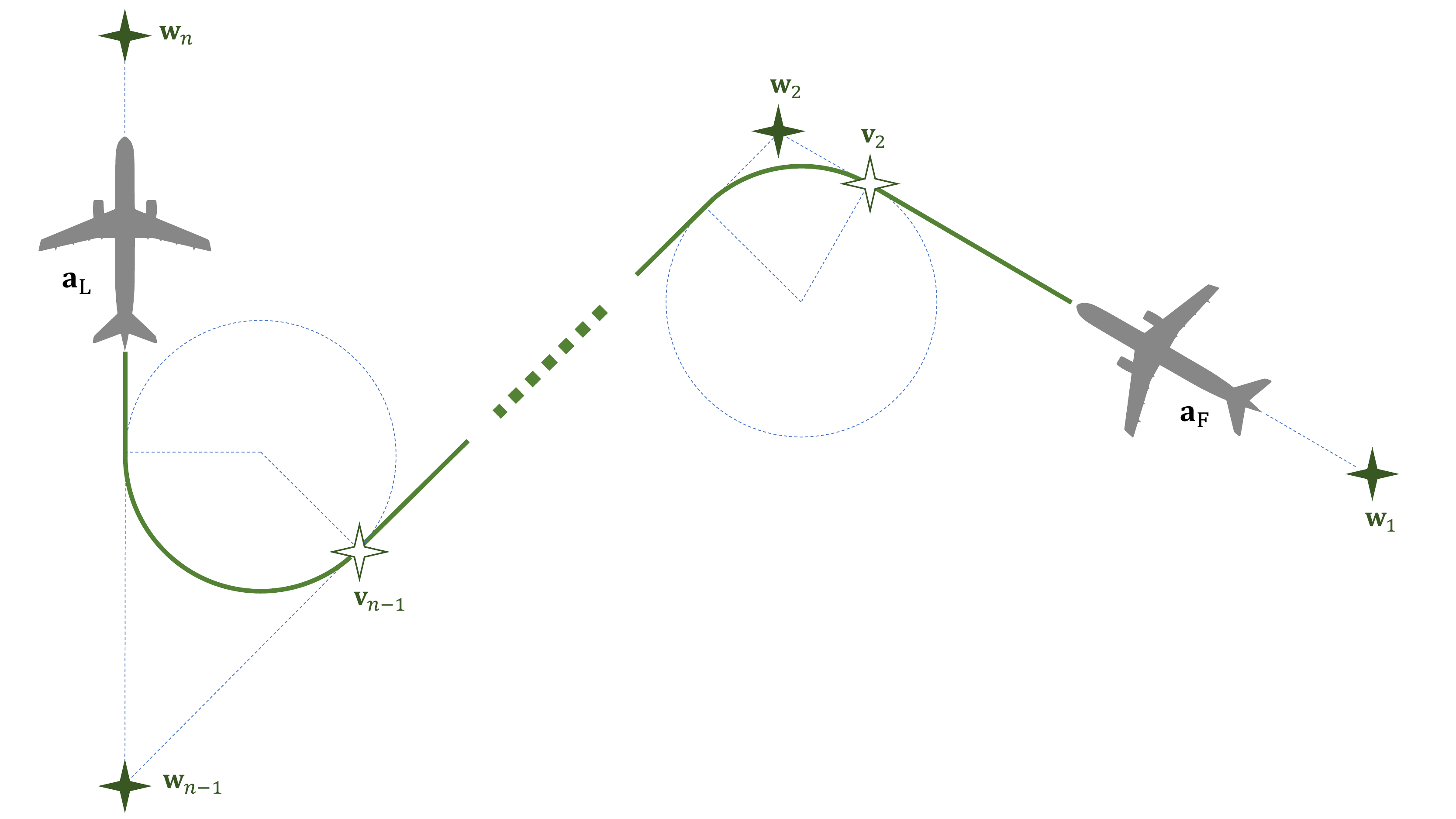}
\caption{Two aircraft describing a Dubins trajectory.}
\label{fig:dubins}
\end{figure}

All the computations are performed by assuming that both the ghost and the missing approach aircraft follow Dubins trajectories \citep{Dubins1957}, since their computation and analysis is relatively simple. Basically, given the initial position and heading of a vehicle, Dubins curves can be used to draw a path in the 2D plane allowing to reach a desired position and heading, on the assumption that the vehicle is moving in a straight line, or using curves with a constant and predefined radius. As an example, Figure \ref{fig:dubins} shows the Dubins trajectory described by two aircraft ($\mathbf{a}_L$, the leader, and $\mathbf{a}_F$, the follower) following the same approach sequence $M=\{\mathbf{w}_1,\mathbf{w}_2,...,\mathbf{w}_{n-1},\mathbf{w}_n\}$. Points $\mathbf{v}_2$ and $\mathbf{v}_{n-1}$ denote transit points between waypoints $\mathbf{w}_2$ and $\mathbf{w}_{n-1}$, and $\mathbf{w}_{n-1}$ and $\mathbf{w}_n$, respectively. Since Dubins curves are not flyable, pilots will ensure that aircraft follow their respective references as closely as possible. More details about this issue can be found in \citep{Casado2021}.

\begin{figure}
\centering
\includegraphics[width=0.8\textwidth]{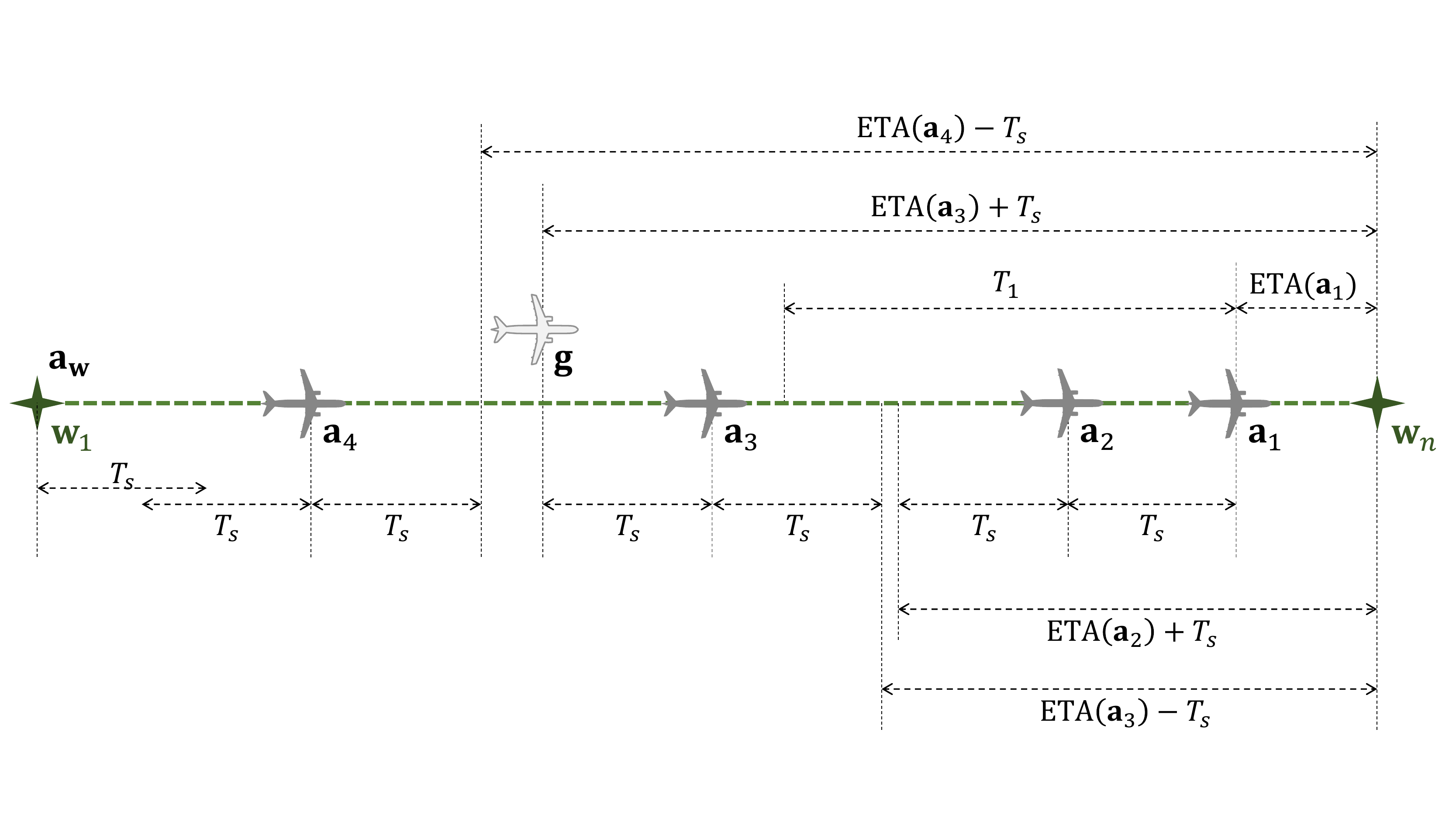}
\caption{Available gap computation.}
\label{fig:gapcomp}
\end{figure}

Dubins curves allow ARS to estimate the time it will take for an approaching aircraft to land (Estimated Time to Arrival, ETA), the time separating two aircraft, and the position of an aircraft within a period of time. All this functionality helps to determine the existence of an available gap in the approach flow. Figure \ref{fig:gapcomp} illustrates this computation. $M=\{\mathbf{w}_1,...,\mathbf{w}_n\}$ is the final portion of an approach sequence, and there is a set of aircraft $\mathbf{a}_1,\mathbf{a}_2,\mathbf{a}_4,\mathbf{a}_4 \in \mathbb{A}$ currently involved in the approach maneuver. At a given time, aircraft $\mathbf{a}_1$ decides to abort the landing. To reinsert it into the approach flow, ARS must find the location of a gap $\mathbf{g} \in \mathbb{A}$ before a threshold time $T_1$ previous to aircraft $\mathbf{a}_1$ (set as the initial search limit), and later than waypoint $\mathbf{w}_1$ (set as the final search limit). In addition, all aircraft maintain a minimum separation time $T_s \in \mathbb{T}$ between them, and the resulting gap (if any) should not be an exception.

\begin{figure}
\centering
\includegraphics[width=0.8\textwidth]{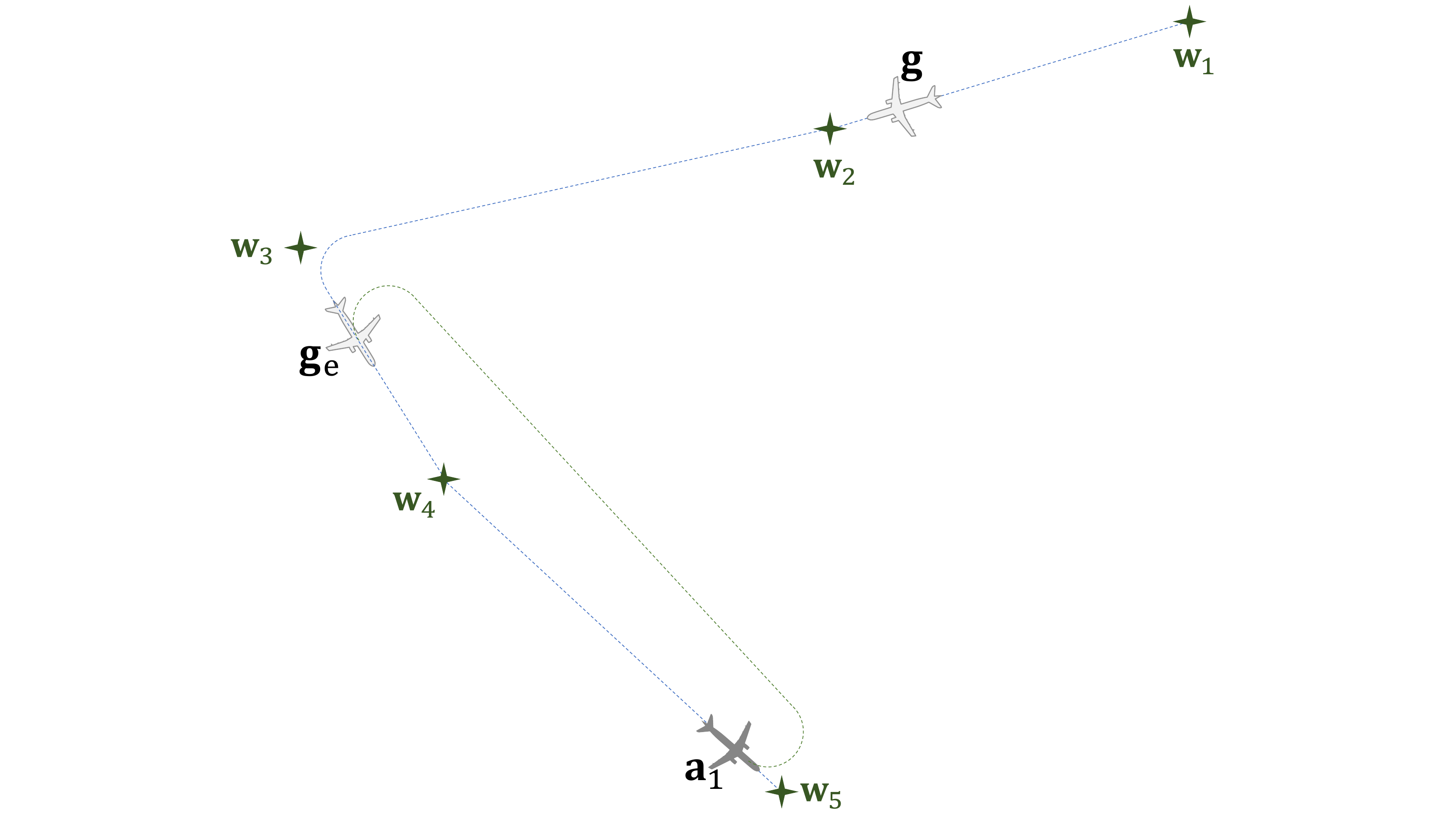}
\caption{Reinjection point computation.}
\label{fig:reinjectionpoint}
\end{figure}

Next, ARS obtains the exact point at which the aircraft missing the approach would intercept a ghost aircraft located in the gap, that is, the reinjection point. Figure \ref{fig:reinjectionpoint} shows an hypothetical situation at Málaga airport, where we assume that a gap for the reinjection of aircraft $\mathbf{a}_1 \in \mathbb{A}$ in the traffic flow (not shown here for clarity) has been found. A ghost aircraft $\mathbf{g} \in \mathbb{A}$ is placed on that gap, and its estimated future position ($\mathbf{g}_e \in \mathbb{A}$) after a certain time is obtained. 
This will be the point used to reinject the aircraft in the approach trajectory.

\begin{figure}
\centering
\includegraphics[width=0.7\textwidth]{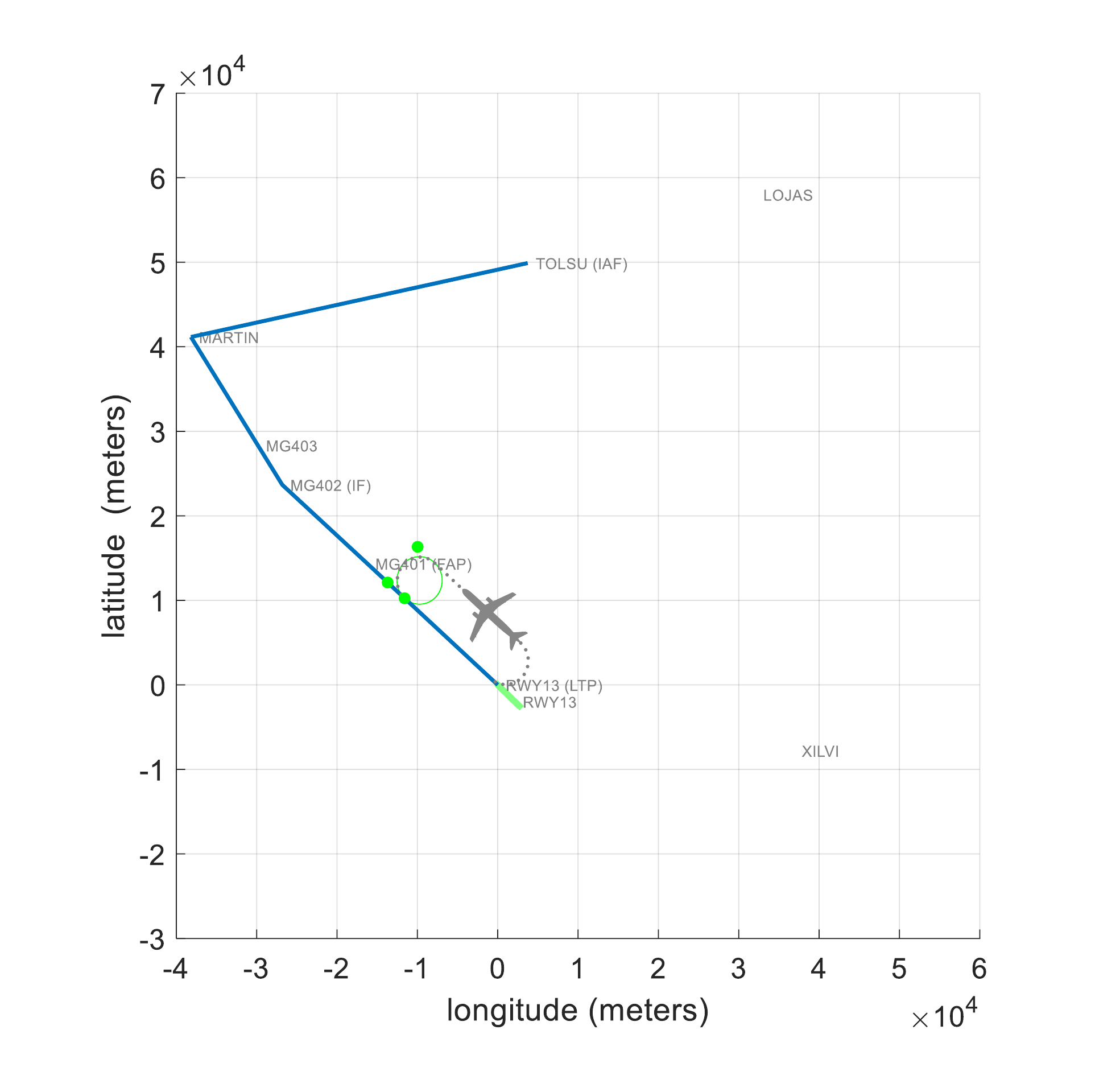}
\caption{Possible reinjection maneuver provided by ARS at Málaga “RWY 13”.}
\label{fig:ARSmalaga}
\end{figure}

Finally, the three new waypoints that the aircraft will use to reach the reinjection point, and that define completely the reinjection maneuver, are computed. Figure \ref{fig:ARSmalaga} shows the trajectory (dotted line) of an aircraft following a missed approach maneuver according to ARS in Málaga airport. Again, for simplicity, we do not show the aircraft flow. Green dots indicate the auxiliary waypoints provided by ARS to guide the aircraft to the reinjection point.

The interested reader can find all the details of a possible implementation of ARS in \citep{Casado2021}.

\section{BADA3 Model}
\label{sec:BADA}


In order to gain awareness of the implications and benefits of the ARS solution presented above, a detailed model for aircraft performance must be provided. In this regard, the Base of Aircraft DAta (BADA) \citep{Eurocontrol2022a} provides a realistic model for determining the performance of any aircraft; in particular, its family 3 aircraft performance models provide a coverage close to 100\% of aircraft types, being considered a reference when attempting to achieve credible modelling of aircraft performances for the nominal part of the aircraft operational envelope. Currently, BADA is widely used in different fields such as research, strategic planning, flow management, ATC ground systems, real time and fast time simulations, and environmental assessments.


For each aircraft analyzed, BADA includes tables for climb, cruise and descent maneuvers. For each maneuver, and for certain altitudes (during an approach would be 0, 500, 1000, 1500, 2000, 3000, 4000, and 6000 ft), the table provides typical values of the behavior of that aircraft (such as horizontal speed, vertical speed, and instantaneous fuel consumption). Yet, neither a traditional missed approach maneuver nor a maneuver aimed at reinjecting the aircraft into the descent flow are comparable to a takeoff situation, as the climb rate is much lower than for the latter. Consequently, in our study, it is not possible to apply the information in this database directly. Instead, we will use the underlying model through which these data were generated, and which is described in this section. Whenever the model requires data for a specific aircraft, we will choose the parameter values corresponding to the Airbus A320.

\subsection{Aircraft Dynamics}

\begin{figure}
\centering
\includegraphics[width=0.5\textwidth]{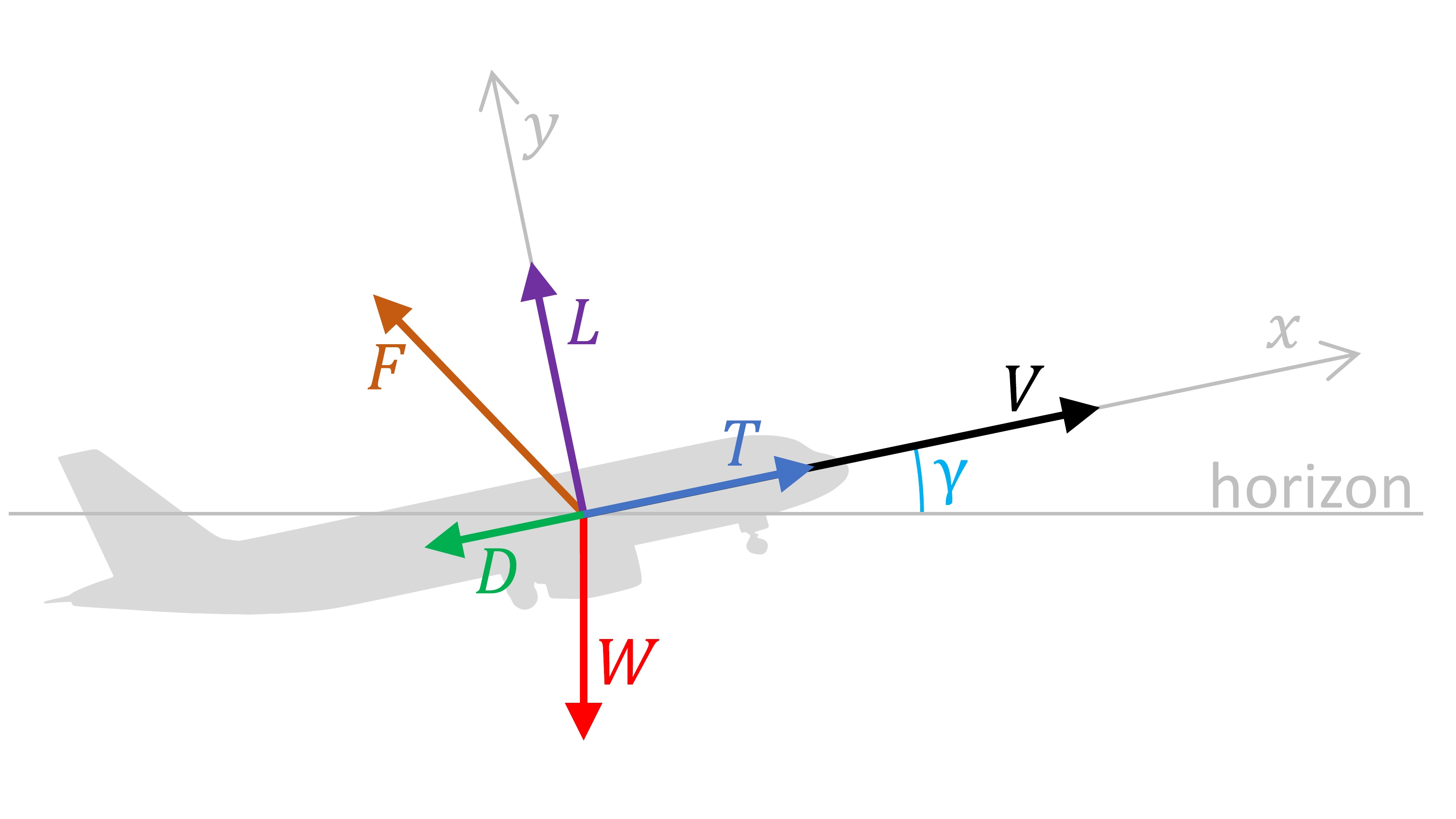}
\caption{BADA3 aircraft model.}
\label{fig:bada3aircraftmodel}
\end{figure}

To gain insight on the different factors affecting the aircraft dynamics, lets assume the situation shown in Figure \ref{fig:bada3aircraftmodel}. The aircraft moves in the direction and speed indicated by vector $V$, maintaining an angle of inclination $\gamma$   with respect to the horizon. We will consider the stability coordinate system (stability axes) where the $x$ axis is associated to the mentioned displacement, and the $y$ axis forms a right angle with the previous one. The movement of the aircraft generates an aerodynamic force that can be broken down into the vertical lift component $L$ (which maintains the aircraft in the air), and the horizontal component of drag $D$ (which acts as a friction force). BADA implements a simplified aerodynamic model in which it is assumed that the engines exert a propulsion force $T$ in the direction of movement (thus maintaining a zero angle of attack). Finally, we denote the weight of the aircraft as $W=mg$. The dynamic model could be decomposed in horizontal forces (Equation \ref{eq:horForces}) and vertical forces (Equation \ref{eq:vertForces}).

\begin{equation}
\sum F_{x} = T - D - W \sin \gamma
\label{eq:horForces}
\end{equation}

\begin{equation}
\sum F_{y} = L - W \cos \gamma
\label{eq:vertForces}
\end{equation}

The advantage of the chosen reference system lies in the fact that the motion is exclusively in the horizontal axis. Applying Newton's second law to Equation \ref{eq:horForces}, and operating, we obtain what is known as Total Energy Model (TEM):

\begin{equation}
\begin{split}
m\dot{V}&=T-D-W\sin\gamma\\
mV\dot{V}&=VT-VD-WV\sin\gamma\\
mV\dot{V}&=VT-VD-W\dot{h}\\
W\dot{h}+mV\dot{V}&=VT-VD\\
mg\dot{h}+mV\dot{V}&=VT-VD\\
mg\dot{h}+mV\dot{V}&=(T-D)V\\
\end{split}
\label{eq:TEM}
\end{equation}

The two terms on the left side of the above expression correspond to the variation of potential and kinetic energy of the aircraft, respectively. The expression indicates that the energy variation is equal to force times velocity.

On the other hand, aerodynamic forces lift $L$ and drag $D$ are defined as

\begin{equation}
L=C_L\frac{1}{2}{\rho}V^2S
\label{eq:eqLift}
\end{equation}

\begin{equation}
D=C_D\frac{1}{2}{\rho}V^2S
\label{eq:eqDrag}
\end{equation}

The ISA (International Standard Atmosphere) model \citep{ISO1975} provides the air density at a certain geopotential height as $\rho=\frac{p}{RT_{air}}$; it is a function of the atmospheric pressure $p$, the air temperature $T_{air}$, and a gas constant that, for air, is equal to $R = 287.05287 m^2/(K s^2)$.
Additionally, according to BADA, an Airbus A320 aircraft has a wing area $S=122.6m^2$.

Taking into account that there is no displacement in the vertical stability axis, by operating Equations \ref{eq:vertForces} and \ref{eq:eqLift} we obtain $C_L\frac{1}{2}{\rho}V^2S=mg\cos\gamma$.
Solving it we are able to determine the lift coefficient as:

\begin{equation}
C_L=\frac{2mg}{{\rho}V^2S}\cos\gamma
\label{eq:eqCL}
\end{equation}

Moreover, the drag coefficient is defined as:

\begin{equation}
C_D=C_{D_0}+C_{D_2}C^2_L
\label{eq:eqCD}
\end{equation}

where $C_{D_0}$ and $C_{D_2}$ are values that depend only on the level of deployment of flaps and gears.
These values are provided by BADA for the different flight modes.
In our case, we have considered the CR (cruise), IC (initial climb) and AP (approaching) modes.
By applying Equations \ref{eq:eqCL} and \ref{eq:eqCD} in Equation \ref{eq:eqDrag}, we obtain the drag force.

Once the drag force $D$ is obtained in Equation \ref{eq:eqDrag}, we can isolate the thrust $T$ in Equation \ref{eq:TEM}, obtaining

\begin{equation}
T=D+m\biggl(\frac{g\dot{h}}{V}+\dot{V}\biggr)
\label{eq:eqT}
\end{equation}

\subsection{Fuel Consumption}
\label{sec:Fuel_consumption}

Based on the different equations describing the interactions between forces during flight, as presented above, we now proceed to derive the equations that allow us to determine the instantaneous fuel consumption; this will ultimately enable determining the overall fuel consumption associated to different maneuvers.  

BADA assumes two modes of fuel consumption, called nominal and minimum, respectively. The nominal consumption model is applied in nearly all occasions. It is proportional to the propulsion force generated by the engines.

\begin{equation}
F_{nom} = {\eta}T
\label{eq:eqFnom}
\end{equation}

Such proportionality depends on the current speed (in knots), and on two constant parameters that BADA provides for each aircraft, $C_{f1}$ and $C_{f2}$, as follows:

\begin{equation}
{\eta}=C_{f1}\biggl(1+\frac{V}{C_{f2}}\biggr)
\label{eq:eq_n}
\end{equation}


In case the aircraft is descending from a certain height (above 2000 ft for the A320 aircraft), it is assumed that the pilot does not want to maintain the horizontal speed (but rather reduce it), so that the engines are in idle thrust mode. In this case, the minimum fuel consumption estimated by BADA depends on the geopotential height, and on two constant parameters (different from the previous ones) that BADA provides for each aircraft, $C_{f3}$ and $C_{f4}$, as follows:

\begin{equation}
F_{min}=C_{f3}\biggl(1-\frac{H_p}{C_{f4}}\biggr)
\label{eq:eqFmin}
\end{equation}

It is also worth mentioning that, in the troposphere (low altitude), the geopotential height $H_p$ can be replaced (with negligible error) by the geometric height $h$ provided by our simulator.

Algorithm \ref{alg:inst-thrust} describes the instantaneous thrust calculation procedure. Initially, air density is obtained using a function implemented by Matlab in its Aerospace blockset (line 3); the lift coefficient is calculated according to Equation \ref{eq:eqCL} (line 4);  after that, we select the appropriate drag coefficients from BADA (line 2) as a function of the current phase of the manoeuvre: ascending (line 6), approaching (line 9), or cruising (line 11); with these coefficients, the drag coefficient is computed (line 14) by applying Equation \ref{eq:eqCD}; then, the drag force is obtained (line 15) applying Equation \ref{eq:eqDrag}; finally, the engine thrust force (line 16) is determined employing Equation \ref{eq:eqT}.

\begin{algorithm}[h!]
\begin{algorithmic}[1]
\caption{Aircraft Instant Thrust model.}
\label{alg:inst-thrust}
\Procedure {InstantThrust}{$V, \dot{V}, \gamma, h, \dot{h}$} 
\LineComment{\textbf{assume} $C_{D0IC}$, $C_{D2IC}$, $C_{D0AP}$, $C_{D2AP}$, $C_{D0CR}$, $C_{D2CR}$}
    \State $\rho = atmosisa(h)$
    \State $C_L = \frac{2mg}{\rho V^2 S}cos(\gamma)$
    \If{$\dot{h} \geq 1$}
        \State $\left\lbrace C_{D0}, C_{D2} \right\rbrace = \left\lbrace C_{D0IC}, C_{D2IC} \right\rbrace$
    \Else
        \If{$\dot{h} \leq -1$}
            \State $\left\lbrace C_{D0}, C_{D2} \right\rbrace = \left\lbrace C_{D0AP}, C_{D2AP} \right\rbrace$
        \Else
            \State $\left\lbrace C_{D0}, C_{D2} \right\rbrace = \left\lbrace C_{D0CR}, C_{D2CR} \right\rbrace$
        \EndIf
    \EndIf
    \State $C_D = C_{D0} + C_{D2} C^{2}_{L}$
    \State $D = C_D \frac{1}{2} \rho V^2 S$
    \State $T = D + m \left( \frac{g \dot{h}}{V} + \dot{V}  \right)$
    \State \Return{$T$}
\EndProcedure
\end{algorithmic}
\end{algorithm}



Concerning Algorithm \ref{alg:inst-fuel}, it describes the instantaneous fuel consumption procedure. 
As stated before, there are two consumption models in BADA. The first one is the standard  model applied in almost all situations; it assumes a fuel consumption (line 7) proportional to the thrust force generated by the engines, according to Equation \ref{eq:eqFnom}. This proportionality (line 6) depends on aircraft speed (line 5) and on two coefficients provided by BADA (line 2) for this particular aircraft. The second option (line 9) employs Equation \ref{eq:eqFmin} to compute idle engine consumption in function of aircraft altitude and on another two coefficients provided by BADA (line 2) for this particular aircraft.

\begin{algorithm}[h!]
\begin{algorithmic}[1]
\caption{Aircraft Instant Fuel Consumption.}
\label{alg:inst-fuel}
\Procedure {InstantFuel}{$h, \dot{h}, T$} 
\LineComment{\textbf{assume} $C_{f1}$, $C_{f2}$, $C_{f3}$, $C_{f4}$}
    \State $h_{ft} = feet(h)$
    \If{$\dot{h} \geq 0$ $\parallel$ $h_{ft} < 2000$  }
        \State $V_{kt} = knots(V)$
        \State $\eta = C_{f1} \left(1 + \frac{V_{kt}}{C_{f2}} \right)/1000$
        \State $F = \eta T$
    \Else
        \State $F = C_{f3} \left( 1 - \frac{h_{ft}}{C_{f4}} \right)$
    \EndIf
    \State \Return{$F$}
\EndProcedure
\end{algorithmic}
\end{algorithm}

Finally, Algorithm \ref{alg:aggregate-fuel} describes the aggregated fuel consumption procedure over time. This algorithm executes an infinite loop (line 3) in which it obtains the current aircraft dynamics parameters (line 4), and computes instant engine thrust (line 5) and fuel consumption (line 6) expressed in Kg/min; after that, it adds to the aggregated fuel value the portion corresponding to a period of time of $1s$ (line 7); then, this value is provided to external instruments (line 8) that may require it; the whole procedure is repeated again after 1 second (line 9).



\begin{algorithm}[h!]
\begin{algorithmic}[1]
\caption{Aircraft Aggregated Fuel Consumption.}
\label{alg:aggregate-fuel}
\Procedure {AggregateFuel}{} 
    \State $F_{a} = 0$
    \While{$True$} 
        \State $get(V, \dot{V}, \gamma, h, \dot{h})$
        \State $T = InstantThrust(V, \dot{V}, \gamma, h, \dot{h})$
        \State $F = InstantFuel(h, \dot{h}, T)$
        \State $F_a = F_a + F/60$
        \State $send(F_a)$
        \State $wait(1)$
    \EndWhile
\EndProcedure
\end{algorithmic}
\end{algorithm}


\section{Performance Evaluation}
\label{sec:perf_eval}

In this section we present the results of our study, which has been  carried out by means of simulation experiments. We start by detailing our simulation tool and how the different experiments were configured, and then proceed to present and discuss the results obtained.

\subsection{Simulation Tool and Experimental Setup}
\label{sec:exp_setup}

Our simulation model has been developed in Matlab/Simulink R2022a \citep{MathWorks2022}, and considers the air traffic flow that approaches to a particular airport runway, including all the elements involved in an approach and landing maneuver.

The tool combines time-based continuous and discrete-event simulation resources available in Simulink. It includes a configurable traffic generator that provides aircraft for the simulation; these appear in the airport airspace, and proceed with the approach and landing procedures according to a programmed chart. We have considered the dynamics and the fuel consumption parameters for the Airbus A320 (see Section \ref{sec:BADA}). Although the tool implements the behavior of both the simulated aircraft following Dubins trajectories, and the real aircraft following flyable trajectories, fuel consumption has been measured for the latter. The ATC, as well as the communications support that allow it to dynamically manage the sequences of waypoints followed by the aircraft, have been also modeled. 

For the experiments of this work, we have considered the approach procedure for the “RWY 13” runway at Málaga airport \citep{ICAO2022b}. We assume that aircraft enter the airport airspace at LOJAS (at 7,000 ft) and they are immediately cleared to the IAF (TOLSU), without executing any holding pattern. Based on ICAO separation standards \citep{ICAO2016}, we have considered five different sequencing patterns, in which approaching aircraft are spaced by $T_s=60 s$, $T_s=90 s$, $T_s=120 s$, $T_s=150 s$, and $T_s=180 s$. We assume that air controllers handle the aircraft sequence following a simple FIFO/FCFS (First-In/Come First-Out/Serve) strategy.

As stated, to reintroduce the aircraft into the descent flow when ARS is in place, a gap must be previously generated. For this, every certain number of aircraft the ATC sequences the following aircraft after $2T_s s$ (instead of after $T_s s$). Finally, we have assumed $T_1=240s$, that is, with ARS an aircraft cannot be reinjected into a gap located less than 4 min away (see Section \ref{sec:ARS}).

\subsection{Results}
\label{sec:results}



\begin{figure}
\centering
\includegraphics[width=0.8\textwidth]{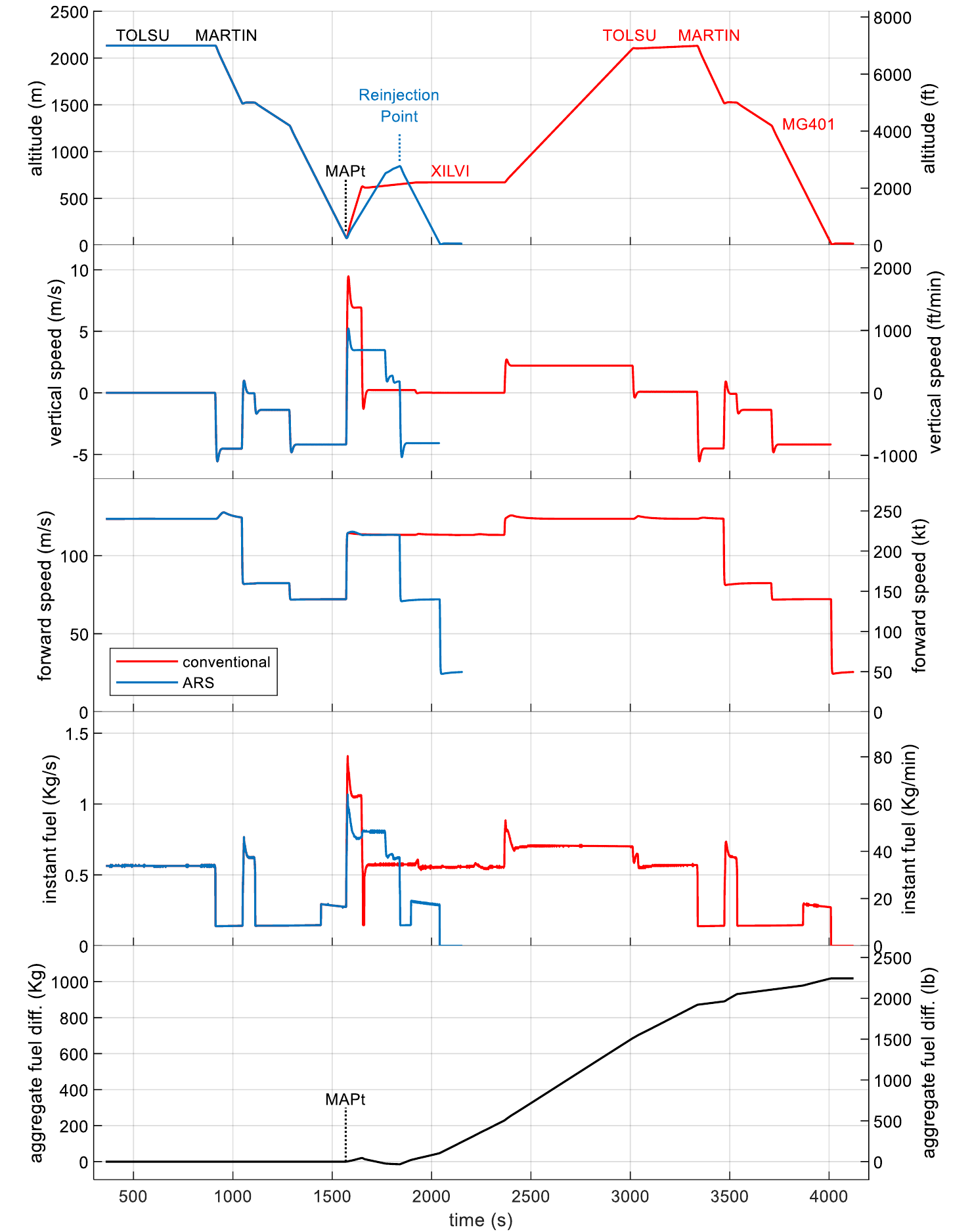}
\caption{Instantaneous comparison between the conventional missed approach procedure and ARS.}
\label{fig:instantfuel}
\end{figure}

Initially we assess the instantaneous fuel consumption for a missed approach situation by comparing the results when an aircraft applies a conventional maneuver, or when the ARS reinjection procedure is adopted.
Aircraft spacing is $T_s=90 s$, and, in the case of ARS, there are 4 aircraft between the one being reinjected and the gap.
Figure \ref{fig:instantfuel} shows the results obtained.

To better understand these results, the first three plots on that figure show, respectively, the aircraft altitude profile, its vertical speed, and its forward speed. Obviously, differences between the series start to appear when the aircraft reaches the MAPt. At this point, we can clearly see an altitude ascend. From this point, in the case of the conventional procedure, we can identify the displacement to XILVI, and a new altitude ascend to again execute the final approach from TOLSU. In the case of ARS, after reaching the established reinjection point, there is a final descend to the runway.

In the second plot in Figure \ref{fig:instantfuel} (vertical speed) we can observe three different constant speed levels that correspond to periods of time during which the aircraft is flying at a constant altitude, descending or ascending. As expected, the second-to-last plot (instant fuel consumption) exhibits the same behavior. In both plots we can see a peak that matches with the sudden aircraft ascent after the MAPt. In the case of the conventional procedure, this peak in speed and fuel consumption is higher because, although the aircraft is climbing up to a lower altitude, it does so in a shorter period of time.

The final plot shows the difference in fuel consumed, but aggregated over time. In this particular case, landing with ARS translates into 1000kg of fuel savings. Note that scales in both the international and aeronautical systems have been included.

\begin{figure}
\centering
\includegraphics[width=0.8\textwidth]{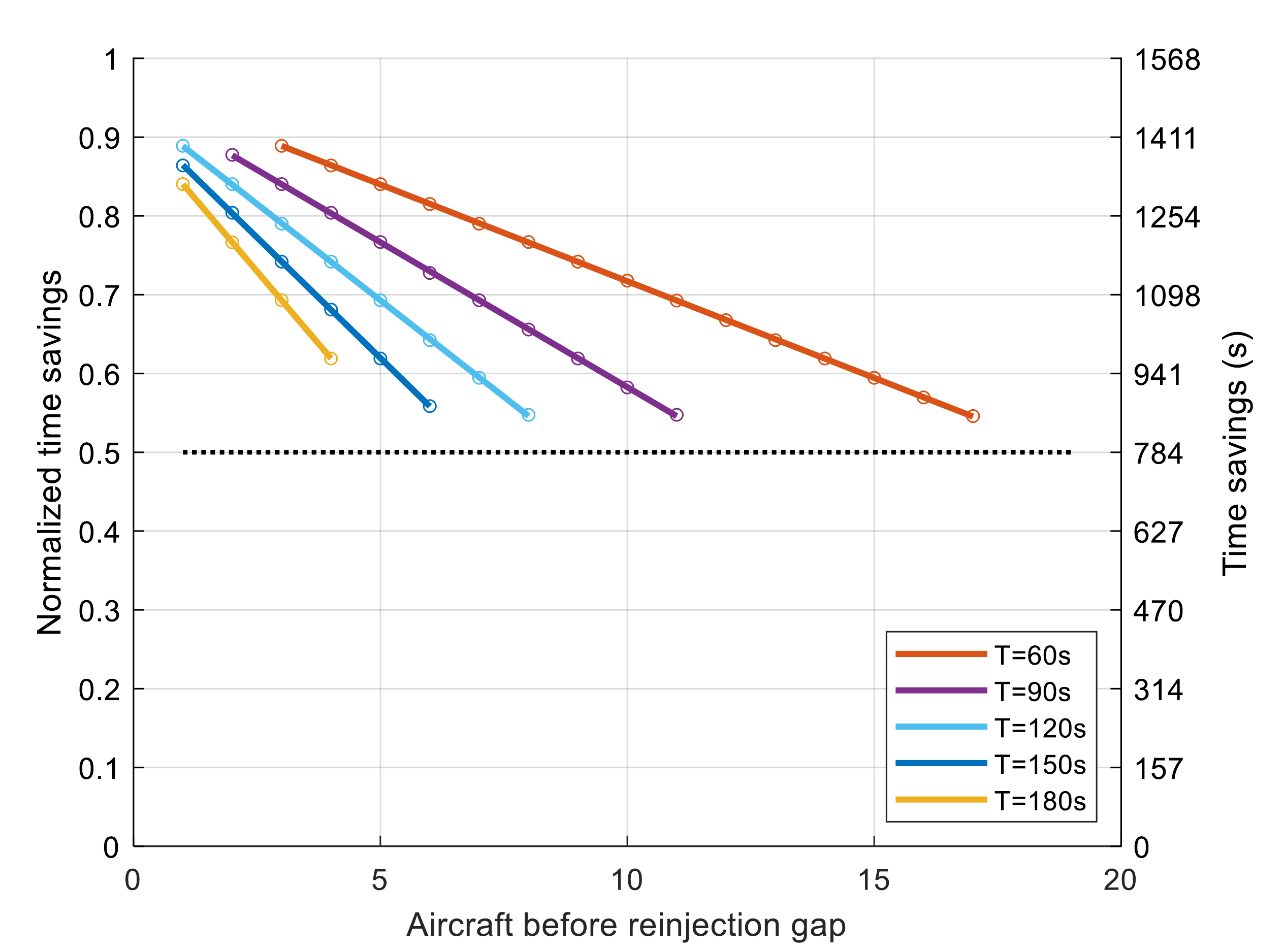}
\caption{Normalized time savings as a function of the distance to the gap.}
\label{fig:timesavings}
\end{figure}

Figure \ref{fig:timesavings} shows the time savings obtained when the ARS is in place at Málaga airport, by taking as a reference the time required by the conventional missed approach procedure. To obtain these results, in each simulation experiment we have measured the time elapsed between the two times when the aircraft is positioned on the MAPt (when it decides to abort the landing, and the when it completes the maneuver successfully). Series indicate sequencing patterns (from 1 to 3 minutes). Horizontal axis represents the amount of aircraft in the approach flow between the one missing the approach and the reinjection gap. Obviously, higher values of this parameter translate into greater distances to the gap. In all cases time savings are observed, which are smaller as the distance to the gap increases, but remaining always above 50\%.

\begin{figure}
\centering
\includegraphics[width=0.8\textwidth]{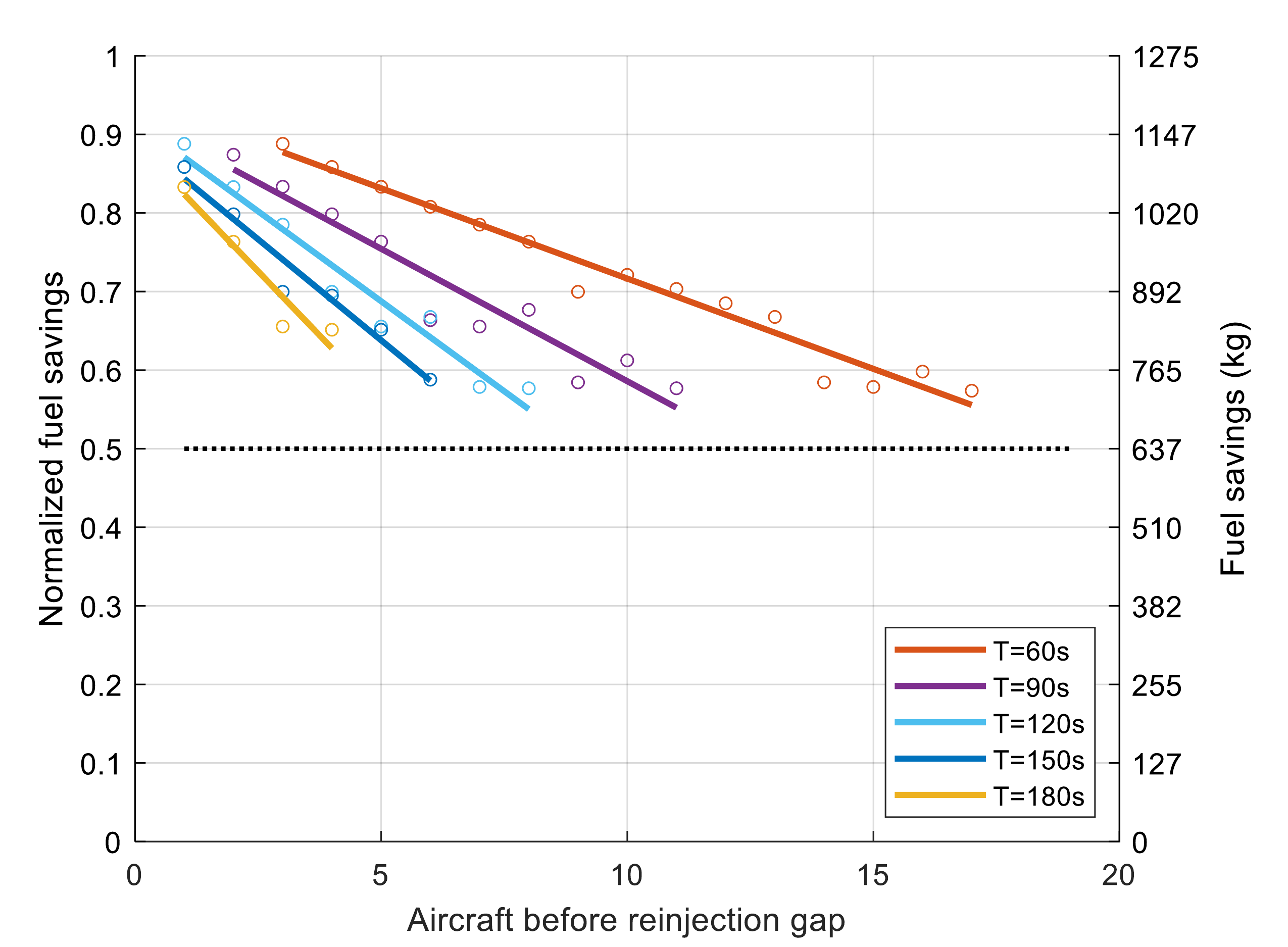}
\caption{Normalized fuel savings as a function of the distance to the gap}
\label{fig:fuelsavings}
\end{figure}

We have also analyzed the fuel consumption savings in the above configurations, based on the fuel consumption model described in section \ref{sec:Fuel_consumption}. Figure \ref{fig:fuelsavings} presents the results obtained. Again, results are normalized with respect to the fuel consumed by the conventional missed approach procedure. As expected, the time savings we have just seen translate perfectly into fuel savings.

\section{Conclusions and Future Works}
\label{sec:conclusions}


From the three phases of a flight (take-off, cruise and landing), it seems that those which imply significant vertical displacements still offer optimizations regarding fuel savings from the aircraft operation point of view. In particular, take-off and landing phases are associated to important engine demands, and these manoeuvres are conditioned by severe safety restrictions. Also, these manoeuvres take place close to airports, affecting populated areas. 

In the case of saturated airports, some studies propose to apply reorganizations of runways and infrastructure in order to reduce fuel consumption, which is a goal not exclusive to our proposal; yet, in general, extensions of these airports is rarely possible due to the implications on neighbouring populated areas.

In a previous work, a new procedure to be used in case of missed approaches was proposed. This procedure could have implications on the safety measures, although the separation between aircraft assigned to the same runway is maintained in any case. 
An analytical model for this new method was developed, and it was validated through simulation techniques.
In this paper we have extended such model to provide an estimation of the fuel savings which could be obtained with the new Aircraft Reinjection System procedure applicable to missed approaches. We prove that a least a 50\% of fuel savings is obtained if compared with the traditional method. This saving is directly related with the CO$_2$ production, and with the reduction with other pollutants.

Continuing with this promising research work, given the immediate benefits it could provide, we will extend the study to evaluate the impact on noise pollution, and also a model of the \emph{ghost} slots used for the aircraft reinjection to evaluate the impact on the whole airport performance.

\section*{Acknowledgments}
This work is derived from R\&D projects RTI2018-098156-B-C52 and RTI2018-096384-B-I00, funded by MCIN/AEI/10.13039/501100011033 and ``ERDF A way of making Europe'', by the Junta de Comunidades de Castilla-La Mancha under grant  SBPLY/19/180501/000159, and by the Universidad de Castilla–La Mancha under grant 2021-GRIN-31042.

\section*{Declarations}

\textbf{Author contribution} The authors confirm contribution to the paper as follows: Conceptualization: R. Casado, A. Bermúdez; Methodology: R. Casado, A. Bermúdez; Software: R. Casado; Validation: M. Carmona, R. Casado, A. Bermúdez, M. Pérez-Francisco, Pablo Boronat, C. Calafate; Formal analysis: M. Carmona, R. Casado, A. Bermúdez, M. Pérez-Francisco, Pablo Boronat; Investigation: M. Carmona, R. Casado, A. Bermúdez; Resources: A. Bermúdez, M. Pérez-Francisco, Pablo Boronat; Data curation: M. Carmona, R. Casado; Writing—original draft: M. Carmona, R. Casado, A. Bermúdez, M. Pérez-Francisco, Pablo Boronat, C. Calafate; Writing—review and editing: R. Casado, A. Bermúdez, M. Pérez-Francisco, Pablo Boronat, C. Calafate; Visualization: M. Carmona, R. Casado, A. Bermúdez, M. Pérez-Francisco, Pablo Boronat, C. Calafate; Supervision: R. Casado, A. Bermúdez, C. Calafate; Project administration: R. Casado, A. Bermúdez, C. Calafate; Funding acquisition: R. Casado, A. Bermúdez, C. Calafate; All authors reviewed the results and approved the final version of the manuscript.

\textbf{Conflict of Interest} On behalf of all authors, the corresponding author states that there is no conflict of interest.


\interlinepenalty=10000
\bibliography{main}



\end{document}